\documentclass[11pt]{article}
\usepackage{epsfig}
\usepackage{graphicx}
\usepackage{epsfig}

\begin{document}

\title
{\bf Structure of Si(114) determined by global optimization methods}%

\author{F.~C.~Chuang$^{1}$, C.~V.~Ciobanu$^{2}$\footnote{Corresponding author.
Email: cciobanu@mines.edu, Phone: 303-384-2119, Fax: 303-273-3602},
C.~Predescu$^3$, C.~Z.~Wang$^1$, and K.~M.~Ho$^1$ \\
$^1$Ames Laboratory -- U.S. Department of Energy and \\Department
of Physics, Iowa State University, Ames, Iowa 50011, USA  \\
$^2$Division of Engineering, Colorado School of Mines, Golden,
Colorado 80401, USA\\
$^{3}$Department of Chemistry and Kenneth S. Pitzer Center for \\
Theoretical Chemistry, University of California, Berkeley, CA
94720, USA}

\date{}

\maketitle

\bigskip
\bigskip
\bigskip
\bigskip

\begin{abstract}

In this article we report the results of global structural
optimization of the Si(114) surface, which is a stable high-index
orientation of silicon. We use two independent procedures recently
developed for the determination of surface reconstructions, the
parallel-tempering Monte Carlo method and the genetic algorithm.
These procedures, coupled with the use of a highly-optimized
interatomic potential for silicon, lead to finding a set of
possible models for Si(114), whose energies are recalculated with
ab-initio density functional methods. The most stable structure
obtained here without experimental input coincides with the
structure determined from scanning tunneling microscopy
experiments and density functional calculations by Erwin, Baski
and Whitman [Phys. Rev. Lett. 77, 687 (1996)].

\end{abstract}

\maketitle \newpage

\section{Introduction}

While the low-index semiconductor surfaces have dominated the
interest of surface scientists for several decades, at present a
considerable amount of work is being dedicated to high-index
orientations. Since the high-index surfaces exhibit more diverse
structural and electronic properties, the adept use of these
properties can constitute the key for various technological
applications, in particular for the fabrication of devices at
length scales where lithographic techniques are not applicable.
Controlling certain physical processes (e.g., growth of
nanostructures in a preferred direction) depends in part on having
knowledge of the structure of the substrate surface. The main
technique for investigating atomic-scale features of surfaces is
scanning tunelling microscopy (STM), although STM alone is only
able to provide "a range of speculative structural models which
are increasingly regarded as solved surface structures"
\cite{woodruff}. A common procedure for finding the
reconstructions of silicon surfaces consists in a combination of
STM imaging and density functional calculations as follows.
Starting from the bulk truncated surface and taking cues from the
experimental data, one proposes several atomic models for the
surface reconstructions. These models are then relaxed using
electronic structure methods; at the end of relaxation, surface
energies and STM images are {\em computed} for each structural
model. A match with the experimental STM data is identified based
on the relaxed lowest-energy structures and their simulated STM
images (e.g., \cite{113dabrowski, science5512, 114-erwin}). As
described, the procedure is heuristic, as one needs to rely
heavily on physical intuition when proposing good candidates for
the lowest energy reconstructions of high-index surfaces.

Treating the reconstruction of semiconductor surfaces as a problem
of global optimization, we have recently developed a parallel
tempering Monte Carlo procedure for studying the structure and
thermodynamics of crystal surfaces \cite{ptmc}, as well as a
genetic algorithm for structure determination \cite{ga}. The use
of such methods can help avoid situations in which the actual
physical reconstruction of a high-index surface is not part of the
set of heuristic models that are considered for computation of
surface energies and comparison with experimental data. Given that
there are examples of semiconductor surfaces (e.g.,
\cite{science5512, mo}) for which the initially proposed models
did not withstand further scientific scrutiny from different
research groups, it appears worthwhile to perform searches for the
structure of some of stable high-index surface orientations of
silicon. One such surface is Si(114), reported to be as stable as
the well studied low-index surfaces Si(001) and Si(111)
\cite{114-erwin}: given this stability of Si(114), it is somewhat
surprising that this surface has not attracted more interest, at
least from a technological perspective. There are few studies of
Si(114) to date, which include the pioneering study  reporting on
the atomic configuration \cite{114-erwin}, and two recent works
reporting on the electronic structure of this surface
\cite{114-uk}.

%
%

Based on scanning tunelling microscope (STM) images combined with
density functional calculations, two atomic models [($2\times 1$)
and $c(c\times 2$)], were proposed for the Si(114) orientation
\cite{114-erwin}. These models have very similar bonding topology,
differing only in terms of dimerization pattern of their surface.
The surface energies of the two models are also similar, as both
can be found on sufficiently large areas of the scanned samples
\cite{114-erwin}. To our knowledge, so far Ref. \cite{114-erwin}
represents the {\em only} proposal for the structure of Si(114).
The purpose of this article is to present several other model
candidates for the structure of Si(114), models that are likely to
be experimentally observed on this surface. Addressing the problem
of atomic structure from a different perspective than the previous
reports \cite{114-erwin, 114-uk}, we perform stochastic searches
for the global minimum configuration of this surface. As we shall
see, the lowest energy configuration (at zero Kelvin) obtained
here from purely theoretical means is consistent with the original
proposal \cite{114-erwin}; however, the global search methods
provide several other structural models with low surface energy,
which could be relevant in various experimental conditions.
%
%
The remainder of this paper is organized as follows. Section II
presents the two recently developed global search algorithms, the
parallel tempering Monte Carlo methods and the genetic algorithm
for structure determination. While brief descriptions of these
methods are provided in Sec. II, we refer the reader to our recent
works \cite{ptmc, ga} for full details related to their
implementation. The results of the structural optimization for
Si(114) surface are presented and discussed in Section III, and
our conclusions are outlined in the last section.

\section{Methods}

\subsection{Parallel-tempering Monte Carlo method} \label{ptmcsection}

The reconstructions of semiconductor surfaces are determined not
only by the efficient bonding of the surface atoms, but also by
the stress created in the process \cite{science5512}. Therefore,
we retain a large number of subsurface atoms when performing a
global search for the lowest energy configuration: this way the
surface stress is intrinsically considered when reconstructions
are sorted out. The number of local minima of the potential energy
is also large, as it scales roughly exponentially
\cite{Sti83,Sti99} with the number of atoms involved in the
reconstruction; by itself, such scaling requires the use of fast
stochastic search methods. One such method is the
parallel-tempering Monte Carlo (PTMC) algorithm
\cite{Gey95,Huk96}, which was shown to successfully find the
reconstructions of a vicinal Si surface when coupled with an
exponential cooling \cite{ptmc}. Before outlining the procedure,
we discuss briefly the computational cell and the empirical
potential used.

The simulation cell (of dimensions $3a \times a \sqrt{2}$ in the
plane of the surface) has a single-face slab geometry with
periodic boundary conditions applied in the plane of the surface,
and no periodicity in the direction normal to it. The ``hot''
atoms from the top part of the slab (10--15~\AA \ thick) are
allowed to move, while the bottom layers of atoms are kept fixed
to simulate the underlying bulk crystal. The area of the
simulation cell and the number of atoms in the cell are kept fixed
during each simulation. Under these conditions, the problem of
finding the most stable reconstruction reduces to the global
minimization of the total potential energy $V(\mathbf{x})$ of the
atoms in the simulation cell (here $\mathbf{x}$ denotes the set of
atomic positions). In terms of atomic interactions, we are
constrained to use empirical potentials because the highly
accurate ab-initio or tight-binding methods are prohibitive as far
as the search itself is concerned. Since this work is aimed at
finding the {\em lowest} energy reconstructions for arbitrary
surfaces, the choice of the empirical potential is important.
After numerical experimentation with several empirical models, we
chose to use the highly optimized empirical potential (HOEP)
recently developed by Lenosky \emph{et al.} \cite{hoep}. HOEP is
fitted to a large database of ab-initio calculations using the
force-matching method, and provides a good description of the
energetics of all atomic coordinations up to $Z=12$.

The parallel tempering Monte Carlo method (also known as the
replica-exchange Monte-Carlo method) consists in running parallel
canonical simulations of many statistically independent replicas
of the system, each at a different temperature $T_1 < T_2 < \ldots
< T_N$. The set of $N$ temperatures $\{T_i,\ i=1,2,...,N \}$ is
called a temperature schedule (or schedule for short). The
probability distributions of the individual replicas are sampled
with the Metropolis algorithm \cite{Met53}, although any other
ergodic strategy can be employed \cite{Hansmann}. Irrespective of
what sampling strategy is being used for each replica, the key
feature of the parallel tempering method is that swaps between
replicas of neighboring temperatures $T_i$ and $T_j$ ($j = i \pm
1$) are proposed and allowed with the conditional probability
\cite{Gey95,Huk96} given by
\begin{equation}
\label{eq:PTMCacc} \min\left\{1, e^{(1/T_j -
1/T_i)\left[V(\mathbf{x}_j)-V(\mathbf{x}_i)\right]/k_B}\right\},
\end{equation}
where $V(\mathbf{x}_i)$ represents the energy of the replica $i$
and $k_B$ is the Boltzmann constant. The conditional probability
(\ref{eq:PTMCacc}) ensures that the detailed balance condition is
satisfied and that the equilibrium distributions are the Boltzmann
ones for each temperature.

In the limit of low temperatures, the PTMC procedure allows for a
geometric temperature schedule \cite{Sug00, Pre03}. To show this,
we note that when the temperature drops to zero, the system is
well approximated by a multidimensional harmonic oscillator, so
the acceptance probability for swaps attempted between two
replicas with temperatures $T < T'$ is given by the incomplete
beta function law \cite{Pre03}
\begin{equation}
\label{eq:AcTT} Ac(T,T') \simeq \frac{2}{B(d/2,d/2)} \int_0^{1/(1
+ R)} \theta^{d/2 - 1}(1 - \theta)^{d/2 -1}d \theta \ ,
\end{equation}
where $d$ denotes the number of degrees of freedom of the system,
$B$ is the Euler beta function, and $R \equiv T' / T$. Since it
depends only on the temperature ratio $R$, the acceptance
probability (\ref{eq:AcTT}) has the same value for any arbitrary
replica running at a temperature $T_i$, provided that its
neighboring upper temperature $T_{i+1}$ is given by
$T_{i+1}=RT_{i}$. The value of $R$ is determined such that the
acceptance probability given by Eq.~(\ref{eq:AcTT}) attains a
prescribed value $p$. Thus, the (optimal) schedule that ensures a
constant probability $p$  for swaps between neighboring
temperatures   is a geometric progression:
\begin{equation}
 T_i = R^{i-1}T_{min},\quad 1 \leq i \leq N,
\label{eq:schedule}
\end{equation}
where $T_{min}=T_1$ is the minimum temperature of the schedule.

The typical Monte Carlo simulation done in this work consists of
two main parts that are equal in terms of computational effort. In
the first stage of the computation, we perform a parallel
tempering run for a range of temperatures  $[T_{min},\ T_{max}]$.
The configurations of minimum energy are retained for each
replica, and used as starting configurations for the second part
of the simulation, in which replicas are cooled down exponentially
until the largest temperature drops below a prescribed value.  As
a key feature of the procedure, the parallel tempering swaps are
not turned off during the cooling steps. Thus, in the second part
of the simulation we are in fact using a combination of parallel
tempering and simulated annealing, rather than a simple cooling.
At the $k$-th cooling step, each temperature from the initial
temperature schedule $\{ T_i, i=1,2,...,N \}$ is decreased by a
factor which is independent of the index $i$ of the replica,
$T_{i}^{(k)} = \alpha_{k} T_{i}^{(k-1)}.$ Because the parallel
tempering swaps are not turned off, we require that at any cooling
step $k$ all $N$ temperatures must be modified by the same factor
$\alpha_{k}$ in order to preserve the original swap acceptance
probabilities. We have used a cooling schedule of the form
\cite{ptmc}
\begin{equation}
\label{eq:cooling} T_{i}^{(k)} = \alpha T_{i}^{(k-1)}=\alpha
^{k-1}T_i \ \ \ \ (k \geq 1),
\end{equation}
where $T_i\equiv T_i^{(1)}$ and $\alpha=0.85$ .

The third and final part of the minimization procedure is a
conjugate-gradient optimization of the last configurations
attained by each replica. The relaxation is necessary because we
aim to classify the reconstructions in a way that does not depend
on temperature, so we compute the surface energy at zero Kelvin
for the relaxed slabs $i,\ i=1,2,..., N$. The surface energy
$\gamma$ is defined as the excess energy (with respect to the
ideal bulk configuration) introduced by the presence of the
surface:
\begin{equation}
\gamma =(E_m-n_m e_b)/A
\end{equation}
where $E_m$ is the potential energy of the $n_m$ atoms that are
allowed to move, $e_b=-4.6124$eV is the bulk cohesion energy given
by HOEP, and $A$ is the surface area of the slab.

\begin{figure}
\begin{center}
 \includegraphics[width=9.0cm]{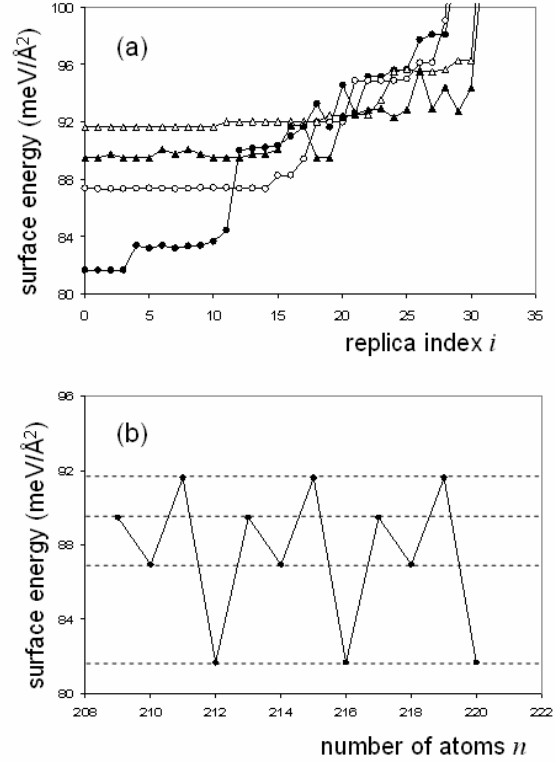}
\caption{(a) Surface energies of the relaxed parallel tempering
replicas $i$, ($0 \leq i \leq31)$ with total number of atoms
$n=216$ (solid circles), 215 (open triangles), 214 (open circles)
and 213 (solid triangles). For clarity, the range of plotted
surface energies was limited from above at 100 meV/\AA$^2$. (b)
Surface energy of the global minimum structure showing a periodic
behavior as a function of $n$, with a period of $\Delta n=4$; this
finding helps narrowing down the set of values for $n$ that need
to be considered for determining the Si(114) reconstructions that
have a $3a\times a\sqrt{2}$ periodic cell.
}%
\label{214_216_ptmc}
\end{center}
\end{figure}

At the end of the simulation, we analyze the energies of the
relaxed replicas. Typical plots showing the surface energies of
the structures retrieved by the PTMC replicas are shown in
Fig.~\ref{214_216_ptmc}(a), for different numbers of particles in
the computational cell. To exhaust all the possibilities for the
numbers of particles corresponding to the supercell dimensions of
$3a \times a \sqrt{2}$, we repeat the PTMC simulation for
different values of $n$ ranging from 208 to 220, and look for a
periodic behavior of the lowest surface energy as a function of
$n$. For the case of Si(114), this periodicity occurs at intervals
of $\Delta n= 4$, as shown in Fig.~\ref{214_216_ptmc}(b).
Therefore, the (correct) number of atoms $n$ at which the lowest
surface energy is attained is $n=216$, up to an integer multiple
of $\Delta n$. As we shall show in the next section, the
repetition of the simulation for different values of $n$ in the
simulation cell can be avoided within a genetic algorithm
approach.

\subsection{Genetic Algorithm}

Like the previous method, the genetic algorithm also circumvents
the intuitive process when proposing candidate models for a given
high-index surface. An advantage of this algorithm over most of
the previous methodologies used for structural optimization is
that the number of atoms involved in the reconstruction, as well
as their most favorable bonding topology, can be found within the
same genetic search. The development of a genetic algorithm (GA)
for surface structure determination was motivated by its
successful application for the structural optimization of atomic
clusters \cite{ga-prl, ga-nature}.

This search procedure is based on the idea of evolutionary
approach in which the members of a generation (pool of models for
the surface) mate with the goal of producing the best specimens,
i.e. lowest energy reconstructions. "Generation zero" is a pool of
$p$ different structures obtained by randomizing the positions of
the topmost atoms (thickness $d$), and by subsequently relaxing
the simulation slabs through a conjugate-gradient procedure. The
evolution from a generation to the next one takes place by mating,
which is achieved by subjecting two randomly picked structures
from the pool to a certain operation (mating) ${\cal
O}$:(A,B)$\longrightarrow$C. The mating operation ${\cal O}$
produces a child structure C from two parent configurations A and
B, as follows. The topmost parts of the parent models A and B
(thickness $d$) are separated from the underlying bulk and
sectioned by an arbitrary plane perpendicular to the surface. The
(upper part of the) child structure C is created by combining the
part of A that lies to the left of the cutting plane and the part
of slab B lying to the right of that plane: the assembly is placed
on a thicker slab, and the resulting structure C is subsequently
relaxed.

A mechanism for the survival of the fittest is implemented as a
defining feature of the genetic evolution. In each generation, a
number of $m$ mating operations are performed. The resulting $m$
children are relaxed and considered for the possible inclusion in
the pool based on their surface energy. If there exists at least
one candidate in the pool that has a higher surface energy than
that of the child considered, then the child structure is included
in the pool. Upon inclusion of the child, the structure with the
highest surface energy is discarded in order to preserve the total
population $p$. As described, the algorithm favors the crowding of
the ecology with identical metastable configurations, which slows
down the evolution towards the global minimum. To avoid the
duplication of members, we retain a new structure only if its
surface energy differs by more than $\delta$ when compared to the
surface energy of any of the current members $p$ of the pool.
Relevant values for the parameters of the algorithm are given in
\cite{ga}: $10\leq p \leq 40$, $m=10$, $d=5$\AA \ , and
$\delta=10^{-5}$meV/\AA$^2$.

\begin{figure}
\begin{center}
 \includegraphics[width=9.0cm]{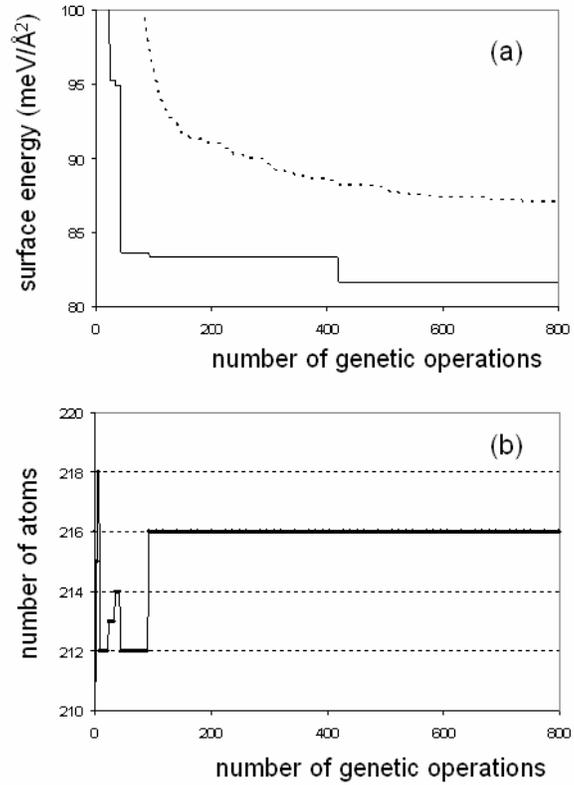}
\caption{(a) Evolution of the lowest surface energy (solid line)
and the average energy (dash line) for a pool of $p=30$ structures
during a genetic algorithm (GA) run with variable $n$ ($210 \leq n
\leq 222$). (b) Evolution of the number of atoms $n$ that
corresponds to the model with the lowest energy from the pool,
during the same GA run. Note that the lowest energy structure of
the pool spends most of its evolution in states with numbers of
atoms that are compatible with the global minimum, i.e. $n=212$
and $n=216$.
}%
\label{ga-var-n}
\end{center}
\end{figure}

We have developed two versions of the algorithm. In the first
version, the number of atoms $n$ is kept the same for every member
of the pool by automatically rejecting child structures that have
different numbers of atoms from their parents (mutants). In the
second version of the algorithm, this restriction is not enforced,
i.e. mutants are allowed to be part of the pool: in this case, the
procedure is able to select the correct number of atoms for the
ground state reconstruction without any increase over the
computational effort required for one single constant-$n$ run. The
results of a variable-$n$ run are shown in Fig.~\ref{ga-var-n}(a)
which shows how the lowest energy and the average energy from a
pool of $p=30$ structures decreases as the genetic algorithm run
proceeds. The plot in Fig.~\ref{ga-var-n}(a) displays typical
features of the evolutionary approach: the most unfavorable
structures are eliminated from the pool rather fast (initial steep
transient region of the graphs) and a longer time is taken for the
algorithm to retrieve the most stable configuration. The lowest
energy structure is retrieved in less than 500 mating operations.
The correct number of atoms [refer to Fig.~\ref{ga-var-n}(b)] is
retrieved much faster, within approximately 100 operations. It is
worth noting that even during the transient period, the
lowest-energy member of the pool spends most of its evolution in a
state with a number of atoms ($n=212$) that is compatible with the
global minimum structure.

The two independent algorithms (PTMC and GA) presented briefly in
this section are able to retrieve a set of possible candidates for
the lowest energy surface structure. We use both of the algorithms
in this work in order to assess how robust their structure
predictions are. As it turns out, the two methods not only find
the same lowest energy structures for each value of the total
number of atoms $n$, but also most of the other low-energy
reconstructions -- a finding that builds confidence in the quality
of the configuration sampling performed here. Since the atomic
interactions are modelled by an empirical potential \cite{hoep},
it is desirable to check the relative stability of different model
structures using higher-level calculations based on density
functional theory; the details of these calculations are presented
next.

\subsection{Density functional calculations} \label{dftsection}

Using the methodologies described above, we build a database of
model structures that are sorted according to the surface energy
given by the HOEP \cite{hoep} interaction model. Since the
empirical potentials may not give a reliable energetic ordering
when a large number of structures are considered, we recalculate
the surface energies of the models in the database at the level of
density functional theory. The calculations where performed with
the plane-wave based PWscf package \cite{pwscf}, using the
Perdew-Zunger \cite{Perdew} exchange-correlation energy. The slab
geometry and the computational parameters are similar to the ones
reported in \cite{114-erwin}; given the increase in computational
speed over the last eight years, we used thicker slabs and a
different sampling of the Brillouin zone. The cutoff for the
plane-wave energy was set to 12 Ry, and the irreducible Brilloiun
zone was sampled with 4 $k$ points. The equilibrium bulk lattice
constant was determined to be $a=$5.41\AA, which was used for all
the surface calculations in this work. The simulation cell has the
single-face slab geometry, with 24 layers of Si, and a vacuum
thickness of 12 \AA. The bottom three layers are kept fixed in
order to simulate the underlying bulk geometry, and the lowest
layer is passivated with hydrogen. The remaining Si layers are
allowed to relax until the forces become smaller than 0.025
eV/\AA.

The surface energy $\gamma$ for each reconstruction is determined
indirectly, by first considering the surface energy $\gamma_0$ of
an unrelaxed bulk truncated slab, then by calculating the
difference $\Delta \gamma = \gamma - \gamma _0 $ between the
surface energy of the actual reconstruction and the surface energy
of bulk truncated slab that has the bottom three layers fixed and
hydrogenated. The energy of the bulk truncated surface, as
computed from a two-faced slab with 24 layers, was found to be
$\gamma_0=143$ meV/\AA$^2$. This indirect procedure for
calculating the surface energies at the DFT level was outlined,
for instance, in Ref. \cite{113dabrowski}.

\section{Structural models for Si(114)}

At the end of the global search procedures, we obtain a set of
model structures which we sort by the number of atoms in the
simulation cell and by their surface energy. Since the empirical
potentials may not be fully transferable to different surface
environments, we study not only the global minima given by the
model for different values of $n$, but also most of the local
minima that are within 15 meV/\AA$^2$ from the lowest energy
configurations. After the global optimizations, the structures
obtained are also relaxed by density functional theory (DFT)
methods as described in Sec.~\ref{dftsection}. The results are
summarized in Table~\ref{table_gamma_smcell}, which will be
discussed next.

\begin{table}
\begin{center}
\begin{tabular}{r c l  l }

\hline \hline
$n$      &  Bond counting         & HOEP    & DFT  \\
\        &  ($db/3a^2\sqrt{2}$)   & (meV/\AA$^2$)& (meV/\AA$^2$)   \\
\hline

216     & 8 & 81.66  &  89.48     \\
\       & 8 & 83.16  &  90.34     \\
\       & 8 & 83.31  &  91.29     \\
\       & 8 & 83.39  &  88.77     \\
\       & 8 & 83.64  &  94.68     \\
\       & 8 & 84.42  &  92.16     \\
\\
215    &  8   & 91.61 &   97.53   \\
\      &  8   & 91.82 &   95.30   \\
\      &  8   & 92.00 &   94.20   \\
\      &  11  & 92.46 &   98.73   \\
\\
214   &  6  & 86.95 &  95.17    \\
\     & 10  & 87.32 &  99.58    \\      
\     & 10  & 87.39 &  98.47    \\
\     & 10  & 87.49 &  93.88    \\
\     & 10  & 88.26 &  95.18    \\
\\
213   &  4  & 89.46  &  90.43   \\
\     &  6  & 89.76  &  94.01   \\
\     &  4  & 90.07  &  90.85   \\
\     &  6  & 91.73  &  94.66   \\
\     &  7  & 93.99  &  90.48   \\
\hline \hline
\end{tabular}
\caption{Surface energies of different  reconstructions for the
Si(114) surface, sorted by the number of atoms $n$ in the $3a
\times a \sqrt{2}$ periodic cell. The second column shows the
number of dangling bonds (counted for structures relaxed with
HOEP) per unit area. The last two columns list the surface
energies given by the HOEP interaction model \cite{hoep} and by
density functional calculations (DFT) \cite{pwscf} with the
parameters described in text. } \label{table_gamma_smcell}
\end{center}
\end{table}

\subsection{Results}
Table 1 lists the density of dangling bonds (db per area), as well
as the surface energies of several different models calculated
using the HOEP potential and DFT. The configurations have been
listed in increasing order of the surface energies computed with
HOEP, as this is the actual outcome of the global optimum
searches. For reasons of space, we limit the number of structures
in the Table~\ref{table_gamma_smcell} to at most six for each
value of the relevant numbers of atoms in the simulation cell.
However, when performing DFT relaxations we consider more
structures than the ones shown in the table because we expect
changes in their energetic ordering at the DFT level. The
inclusion of a larger number of structures helps avoid excessive
reliance on the empirical potential \cite{hoep}, which is mainly
used a fast way to provide physically relevant reconstructions
(i.e. where each atom at the surface has at most one dangling
bond).

\begin{figure}
\begin{center}
 \includegraphics[width=12.0cm]{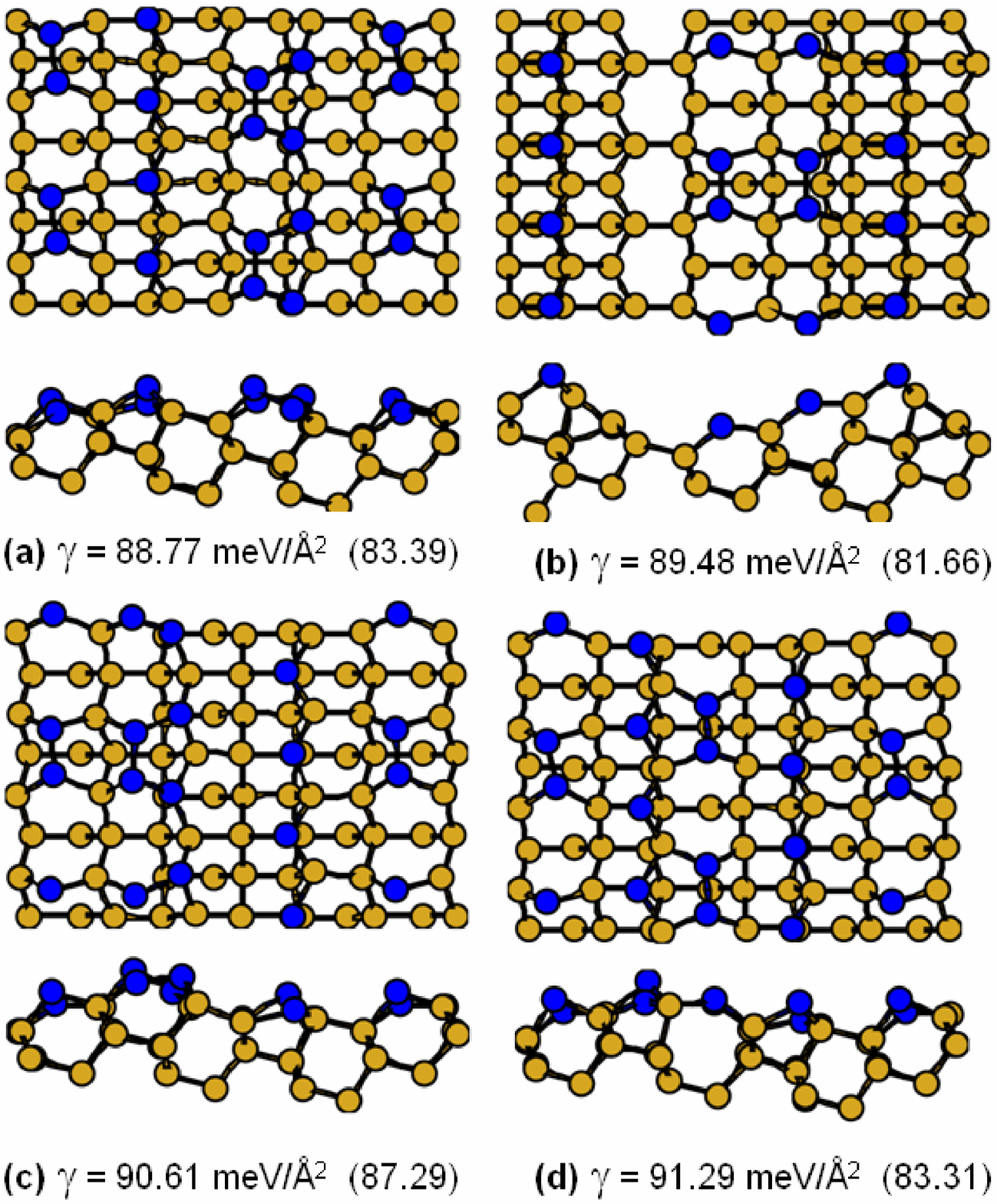}
\caption{Structural models (top and side views) of
Si(114)-(2$\times 1$), with $n=216$ atoms per unit cell after
relaxation with density functional methods \cite{pwscf}. The
surface energy $\gamma$ computed from first-principles is
indicated for each structure, along with the corresponding value
(in parentheses) obtained using the empirical potential
\cite{hoep}. The darker shade marks the undercoordinated atoms.
}%
\label{216-si114}
\end{center}
\end{figure}

\begin{figure}
\begin{center}
 \includegraphics[width=12.0cm]{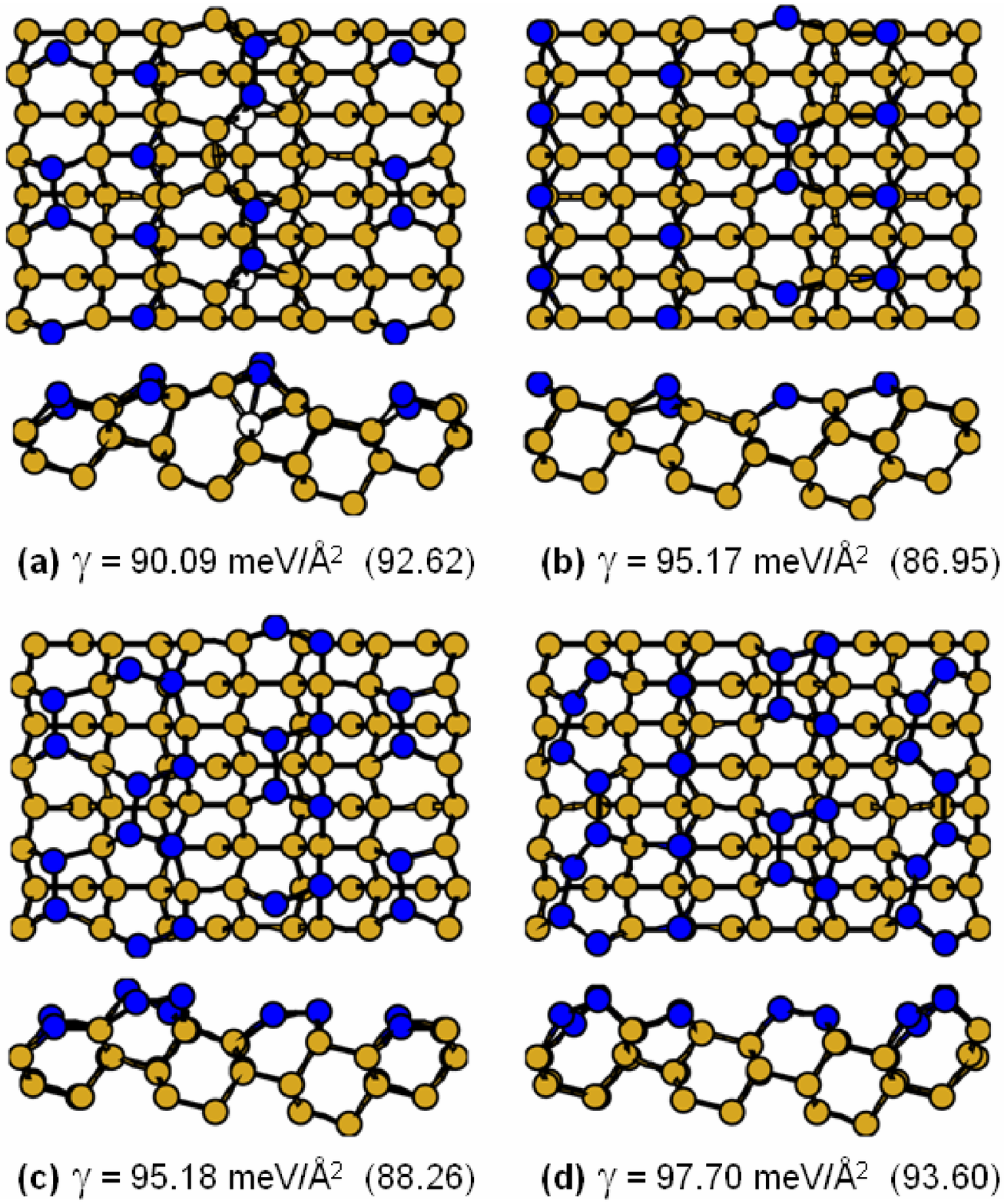}
\caption{Structural models (top and side views) of
Si(114)-(2$\times 1$), with $n=214$ atoms per unit cell after
relaxation with density functional methods \cite{pwscf}. The
surface energy $\gamma$ computed from first-principles is
indicated for each structure, along with the corresponding value
(in parentheses) obtained using the empirical potential
\cite{hoep}. The darker shade marks the undercoordinated atoms,
while the overcoordinated atoms are shown in white.
}%
\label{214-si114}
\end{center}
\end{figure}

\begin{figure}
\begin{center}
 \includegraphics[width=12.0cm]{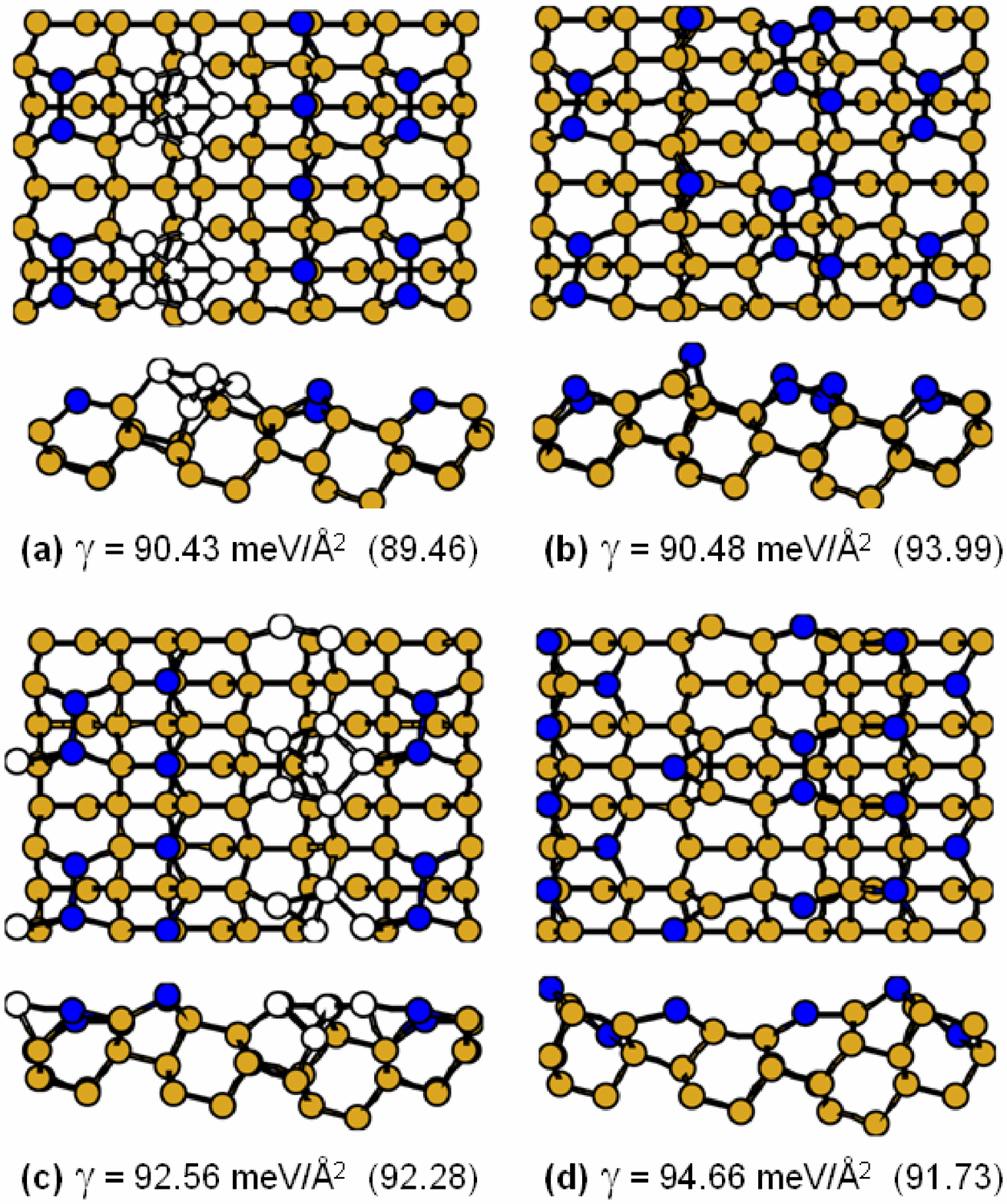}
\caption{Structural models (top and side views) of
Si(114)-(2$\times 1$), with $n=213$ atoms per unit cell after
relaxation with density functional methods \cite{pwscf}. The
surface energy $\gamma$ computed from first-principles is
indicated for each structure, along with the corresponding value
(in parentheses) obtained using the empirical potential
\cite{hoep}. The darker shade marks the undercoordinated atoms,
while the white atoms are either four-coordinated or
overcoordinated.
}%
\label{213-si114}
\end{center}
\end{figure}

Table \ref{table_gamma_smcell} and Fig.~\ref{214_216_ptmc}(b)
suggest that the most unfavorable number of atoms in the
simulation cell is $n=215$, both at the level of HOEP and at the
level of DFT. Therefore, it is justifiable to focus our attention
on the other three values of $n$ ($n=216,\ 214$ and 213), which
yield considerably lower surface energies. For each of these
numbers of atoms, we present four low energy structures (as given
by DFT), which are shown in Figs.
\ref{216-si114}--\ref{213-si114}. These structures are not
necessarily the same as those enumerated in
Table~\ref{table_gamma_smcell}, as they are chosen based on their
DFT surface energies. Since the global optimization has not been
performed at the DFT level, the reader could argue that the lowest
energy structure obtained after the sorting of the DFT-relaxed
models may not be the DFT global minimum. While we found that a
{\em thorough} sampling for systems with $\sim 200$ atoms is
impractical at the DFT level, we have performed DFT relaxations
for most of the local minima given by HOEP. Therefore, given the
rather large set of structural candidates with different
topological features considered here, the possibility of missing
the actual reconstruction for Si(114) is much diminished in
comparison with heuristic approaches.

We will now describe in turn the surface models corresponding to
$n=216$, 214 and 213. After the DFT relaxation, the lowest energy
model that we found has turned out to be the same as the one
proposed by Erwin and coworkers \cite{114-erwin}, perhaps with the
exception of different relative tilting of the surface bonds. The
model is shown in Fig.~\ref{216-si114}(a), and it is characterized
by the presence of dimers, rebonded atoms and tetramers occurring
in this order along the (positive) [22${\overline 1}$] direction.
These features have been well studied \cite{114-erwin, 114-uk},
and we shall not insist on them here. The surface energy of the
most stable model for Si(114)-($2\time 1$) reconstruction is
$\gamma = 88.77$ meV/\AA$^2$. Although this surface energy is
somewhat different from the previously reported value of $85$
meV/\AA$^2$ \cite{114-erwin}, the discrepancy between these
absolute values can be attributed to the somewhat different
computational parameters (slab thickness, number of $k$ points)
and/or different pseudopotentials.

It is notable that a different succession of the above-mentioned
atomic scale features is also characterized by a low surface
energy: specifically, dimers, tetramers and rebonded atoms (in
this order along [22${\overline 1}$]), as shown in
Fig.~\ref{216-si114}(c), give a surface energy which is only
$\sim$2 meV/\AA$^2$ higher than that of the Erwin {\em et al.}
model shown in \ref{216-si114}(a).  This surface energy gap is
apparently large enough to allow for another configuration [see
Fig.~\ref{216-si114}(b)] with a surface energy that lies between
the values corresponding to the first two models described above.
As shown in Fig.~\ref{216-si114}(b), this new model has two
consecutive dimer rows followed by a row of rebonded atoms, and
arrangement that  gives rise to surface corrugations of 0.4--0.5
nm. Remarkably, this corrugated model [Fig.~\ref{216-si114}(b)] is
almost degenerate with the planar, $(2\times 1)$ structure shown
in Fig.~\ref{216-si114}(a). The last panel of Fig.~\ref{216-si114}
shows another planar model of Si(114), made of dimers, rebonded
atoms and inverted tetramers, with the latter topological feature
distinguishable as a seven-member ring when viewed along the [110]
direction.

Dimers, rebonded atoms and tetramers also occur on low-energy
structural models with $n=214$, as shown in Fig.~\ref{214-si114}.
The most favorable structure with $n=214$ that we found [depicted
in Fig.~\ref{214-si114}(a)] has a 5-coordinated subsurface atom
and a 4-coordinated surface atom per unit cell. These topological
features are determined by the bonding of a subsurface atom with
one of the atoms of a tilted surface dimer; the corresponding
surface energy is 90.09 meV/\AA$^2$. Other structures with $n=214$
atoms [examples shown in Fig.~\ref{214-si114}(b)--(d)] generally
have higher energies than models with $n=216$, (refer to
Table~\ref{table_gamma_smcell}) most likely because the two
missing atoms lead to pronounced strains in the surface bonds.

The analysis of simulation slabs with $n=213$ atoms reveals novel
atomic scale features. Energetically favorable configurations with
$n=213$ [\ref{213-si114}(a) and (c)] show a 5-atom ring on the
surface stabilized by a subsurface interstitial, a structural
complex that was first encountered in the case of Si(113) surface
\cite{113dabrowski}. Structures in Figs.~\ref{213-si114}(a) and
(c) differ in terms of the succession of the topological features
along the [22${\overline 1}$] direction, i.e. dimers, 5-member
rings, rebonded atoms [\ref{213-si114}(a)] as opposed to dimers,
rebonded atoms and 5-member rings [\ref{213-si114}(c)]. The model
in Fig.~\ref{213-si114}(a) is degenerate with the one shown in
\ref{213-si114}(b), as their relative surface energy is much
smaller than the 1--2 meV/\AA$^2$ expected accuracy of the
relative surface energies determined here. The reconstruction
\ref{213-si114}(b) is very similar to the lowest energy structure
in Fig.~\ref{216-si114}(a) (achievable with $n=212$): the only
different feature is the extra atom lying in between two rebonded
atoms and sticking out of the surface [refer to
Fig.~\ref{213-si114}(b)]. Likewise, the model in
Fig.~\ref{213-si114}(d) can be obtained from structure
\ref{216-si114}(b) by adding one atom per unit cell in such a way
that it bridges the two atoms of a dimer on one side, and rebonds
on the other side.

\subsection{Discussion}

The data in Table~\ref{table_gamma_smcell} shows clearly that the
density of dangling bonds at the Si(114) surface is, in fact
uncorrelated with the surface energy. The lowest number of dbs per
area reported here is 4, and it corresponds to $n=213$ and
$\gamma=90.43$ meV/\AA$^2$ at the DFT level. The optimum structure
[\ref{216-si114}(a)] , however, has twice as many dangling bonds
but the surface energy is smaller, $88.77$ meV/\AA$^2$.
Furthermore, for the same number of atoms in the supercell
($n=216$) and the same dangling bond density ($8db/3a^2\sqrt{2}$),
the different reconstructions obtained via global searches span an
energy interval of at least 5 meV/\AA$^2$. These findings
constitute a clear example that the number of dangling bonds can
not be used as a criterion for selecting model reconstructions for
Si(114); we expect this conclusion to hold for many other
high-index semiconductor surfaces as well.

The HOEP surface energy and the DFT surface energy also show very
little correlation, indicating that the transferability of the
interaction model \cite{hoep} for Si(114) is not as good as, for
instance, in the case of Si(001) and Si(105) \cite{ptmc}. The most
that can be asked from this model potential \cite{hoep} is that
the observed reconstruction \cite{114-erwin} is amongst the lower
lying energetic configurations --which, in this case it is. We
have also tested the transferability of HOEP for the case of
Si(113), and found that, although the ad-atom interstitial models
\cite{113dabrowski} are not the most stable structures, they are
still retrieved by HOEP as local minima of the surface energy. We
found that the low-index (but much more complex) Si(111)-$(7\times
7)$ reconstruction is also a local minimum of the HOEP interaction
model, albeit with a very high surface energy. Other tests
indicated that, while the transferability of HOEP to the Si(114)
orientation is marginal in terms of sorting structural models by
their surface energy, this potential \cite{hoep} performs much
better than the more popular interaction models \cite{sw,
tersoff}, which sometimes do not retrieve the correct
reconstructions even as local minima. Therefore, HOEP is very
useful as a way to find different local minimum configurations for
further optimization at the level of electronic structure
calculations.

A practical issue that arises when carrying out the global
searches for surface reconstructions is the two-dimensional
periodicity of the computational slab. In general, if a periodic
surface pattern has been observed, then the lengths and directions
of the surface unit vectors may be determined accurately through
experimental means (e.g., STM or LEED analysis): in those cases,
the periodic vectors of the simulation slab should simply be
chosen the same as the ones found in experiments. When the surface
does not have two-dimensional periodicity, or when experimental
data is difficult to analyze, then one should systematically test
computational cells with periodic vectors that are integer
multiples of the unit vectors of the bulk truncated surface, which
are easily computed from knowledge of crystal structure and
surface orientation. There is no preset criterion as to when the
incremental testing of the size of the surface cell should be
stopped --other than the limitation imposed by finite
computational resources; nevertheless, this approach gives a
systematic way of ranking the surface energies of slabs of
different areas, and eventually finding the global minimum surface
structure.

Motivated by a previous finding that larger unit cells can lead to
models with very low surface energies (see, for instance the
example of Si(105) \cite{ptmc, ga}), we have also performed global
minimum search using GA for slabs of dimensions $6a\times
a\sqrt{2}$, which correspond to a doubling of the unit cell in the
[22${\overline 1}$] direction. The ground state structure at the
HOEP level found in this case is still the corrugated model
\ref{216-si114}(b) with a surface energy of $\gamma=81.66$
meV/\AA$^2$. As a low-lying configuration we again retrieve the
original model \cite{114-erwin} with $\gamma=83.39$ meV/\AA$^2$.
Furthermore, we also find a several models that have surface
energy in between the two values, characterized by the presence of
different $3a\times a\sqrt{2}$ structures in the two halves of the
$6a \times a\sqrt{2}$ simulation cell. This finding suggests that
[1$\overline{1}$0]-oriented boundaries between different $3a\times
a\sqrt{2}$ models on Si(114) are not energetically very costly:
this is consistent with the experimental reports of Erwin {\em et
al.}, who indeed found $(2\times 1)$ and $c(2\times 2)$ structures
next to one another \cite{114-erwin}.

\section{Concluding remarks}

In conclusion, we have obtained and classified structural
candidates for the Si(114) surface reconstructions using global
optimization methods and density functional calculations. We have
used both parallel-tempering Monte Carlo procedure coupled with an
exponential cooling \cite{ptmc}, as well as the genetic algorithm
\cite{ga}. Both of the methods are used in conjunction with the
latest empirical potential for silicon \cite{hoep}, which has a
better transferability in comparison with more popular potentials
\cite{sw, tersoff}. We have built a large database of structures
(reported, in part, in Table~\ref{table_gamma_smcell}) which were
further optimized at the DFT level. The lowest energy structure
that we found (Fig.~\ref{216-si114}(a)) after the DFT relaxation
is the same as the one originally reported for Si(114)-($2 \times
1$) in \cite{114-erwin}.

In addition, we have discovered several other types of structures
(refer to Figs.~\ref{216-si114}(b), \ref{216-si114}(c) and
\ref{214-si114}(a) and \ref{213-si114}(a)) that are separated
(energetically) by 1--2 meV/\AA$^2$ from the lowest energy model
\cite{114-erwin}. Given that the relative surface energies at the
DFT level have an error of $\pm 1$ meV/\AA$^2$, and that the
experiments of Erwin {\em et al.} \cite{114-erwin} already
identified two reconstructions ([($2\times 1$) and $c(2\times 2$)]
whose surface energies are within 1--2 meV/\AA$^2$ from one
another, it is conceivable that some of the models in Figs.
\ref{216-si114}--\ref{213-si114} could also be found on the
Si(114) surface. This prediction could be tested, e.g., by
high-resolution transmission electron microscopy experiments such
as the ones reported recently for the Si(5512) surface
\cite{5512takeguchi}. Low-energy electron diffraction experiments,
as well as more STM measurements could also shed light on whether
there exist other structural models on a clean Si(114) surface
than initially reported in Ref.~\cite{114-erwin}.

{\bf Acknowledgments.} Ames Laboratory is operated for the U.S.
Department of Energy by Iowa State University under Contract No.
W-7405-Eng-82; this work was supported by the Director of Energy
Research, Office of Basic Energy Sciences. This work has also been
supported in part by National Science Foundation Grants No.
CHE-0095053 and CHE-0131114. The genetic algorithm runs were
performed at the National Energy Research Supercomputing Center
(NERSC) in Berkeley, California. We gratefully acknowledge the use
of the EMSL computational resources at the Pacific Northwest
National Laboratory, where the density functional calculations
were completed. The Monte Carlo simulations were done at the
Center of Advanced Scientific Computation and Visualization
(CASCV) at Brown University; we thank Professor J. D. Doll for
generously allowing us to use his computational facilities at
CASCV.

\end{document}